\documentclass[twocolumn]{aastex631}

\usepackage{amsmath}
\usepackage{leftidx}
\usepackage{graphicx}
\usepackage{subfigure}
\usepackage{paralist}

\shorttitle{Xu article}
\shortauthors{Xu et al.}

\graphicspath{{./}{figures/}}

\begin{document}

\title{The Impact of Inelastic Collisions with Hydrogen on NLTE Copper Abundances in Metal-Poor Stars}

\correspondingauthor{Jianrong Shi}
\email{sjr@bao.ac.cn}

\author{Xiaodong Xu}
\affil{Beijing Planetarium, Beijing Academy of Science and Technology, Beijing 100044, PR China}

\author{Jianrong Shi}
\affil{Key Laboratory of Optical Astronomy, National Astronomical Observatories, Chinese Academy 
of Science, Beijing 100012, PR China}
\affil{University of Chinese Academy of Sciences, Beijing 100049, PR China}

\author{Xiaofeng Wang}
\affil{Beijing Planetarium, Beijing Academy of Science and Technology, Beijing 100044, PR China}
\affil{Physics Department and Tsinghua Center for Astrophysics (THCA), Tsinghua University, Beijing 100084, PR China}

\begin{abstract}
We investigate the non-local thermodynamic equilibrium (NLTE) analysis for \ion{Cu}{1} lines
with the updated model atom that includes quantum-mechanical rate coefficients of Cu\,$+$\,H and Cu$^+$\,$+$\,H$^-$ inelastic 
collisions from the recent study of \cite{2021MNRAS.501.4968B}. The influence of these data on NLTE abundance determinations
has been performed for six metal-poor stars in a metallicity range of $-$2.59\,dex$\,\le$\,[Fe/H]\,$\le$\,$-$0.95\,dex. For \ion{Cu}{1} lines, 
the application of accurate atomic data leads to a decrease in the departure from LTE and lower copper abundances
compared to that obtained with the Drawin's theoretical approximation. To verify our adopted copper atomic model,
we also derived the LTE copper abundances of \ion{Cu}{2} lines for the sample stars. A consistent copper abundance 
from the \ion{Cu}{1} (NLTE) and \ion{Cu}{2} (LTE) lines has been obtained, which indicates the
reliability of our copper atomic model. It is noted that the [Cu/Fe] ratios increase with increasing metallicity
when $\sim$\,$-$2.0\,dex\,$<$\,[Fe/H]\,$<$\,$\sim$\,$-$1.0\,dex, favoring a secondary (metallicity-dependent) copper production.
\end{abstract}


\section{Introduction} \label{sec:intro}

The investigation of the chemical compositions of stars encoded in stellar spectra is one of the fundamental problems in contemporary
astrophysics \citep[e.g.][]{1994ApJS...91..749M, 2002ARA&A..40..487F, 2005ARA&A..43..481A, 2011MNRAS.414.3231K, 2014IAUS..298..355M, 2016A&ARv..24....9B}. 
Non-local thermodynamic equilibrium (NLTE) modeling of stellar spectra
plays an important role in deriving both relative and absolute abundances for various chemical elements in stars, particularly in the
case of metal-deficient objects. Contrary to LTE assumptions, NLTE modeling requires lots of information about the radiative and
collision inelastic processes in collision with electrons and hydrogens. For a given atomic species, the mutual neutralization processes involving the
collision of positive ions with hydrogen negative ions are also necessary. The main uncertainty for NLTE studies has often been introduced,
due to the lack of reliable collisional data with hydrogen particles \citep{2005ARA&A..43..481A, 2016A&ARv..24....9B}.

Copper is an interesting iron-peak element, which is theoretically thought to be synthesized in multiple nucleosynthetic
processes \citep{2004MmSAI..75..741B}: 
\begin{inparaenum}[\itshape a\upshape)] 
\item the weak \emph{s}-process \citep[i.e. occurring in massive stars during core-helium, carbon-shell, and 
the explosive complete Ne burning stages,][]{1995ApJS..101..181W,2003ApJ...592..404L,2010ApJ...710.1557P}, or the main \emph{s}-process 
\citep[i.e. taking place in low and intermediate-mass asymptotic giant branch stars,][]{1999ApJ...525..886A}; 
\item the explosive nucleosynthesis, either in long-lived Type Ia supernovae
 \citep{1993A&A...272..421M,1999ApJS..125..439I,2004A&A...425.1029T,2014MNRAS.438.1762F},
or in Type II supernovae \citep{1995ApJS...98..617T};
\item the \emph{weak} \emph{sr}-process \citep[i.e. operating during core He-burning and during the subsequent convective shell C-burning
phase in massive stars,][]{2004MmSAI..75..741B}. 
\end{inparaenum}
However, there is no consensus on the respective importance of the above processes. For instance,
\cite{1993A&A...272..421M}, \cite{2002A&A...396..189M} and \cite{2003AJ....125.2018S} supported the point that the main source of copper was SNe\,Ia.
While \cite{2004MmSAI..75..741B} suggested that most Cu was synthesized by the \emph{weak} \emph{sr}-process in massive stars, which was
subsequently supported by the predictions of the Galactic chemical evolution (GCE) model of \cite{2007MNRAS.378L..59R} and \cite{2010A&A...522A..32R}. 
Therefore, detailed and reliable observations of Cu abundances can help to figure out the problem.

So far, most analyses of the copper abundances in stellar atmospheres adopted the LTE assumption. \cite{2014ApJ...782...80S} for the first time modeled Cu
in the NLTE framework and pointed out that the important NLTE mechanism influencing the \ion{Cu}{1} spectra was ultraviolet (UV) overionization and 
the NLTE effects were essential for deriving Cu abundances, which was later confirmed by \cite{2015ApJ...802...36Y, 2016A&A...585A.102Y}. \cite{2018MNRAS.473.3377A} developed
their own Cu atomic model, and determined the NLTE copper abundances within a wide metallicity range, -4.2\,<\,[Fe/H]\,<\,-1.4. They found
that the NLTE effects were strong for copper in metal-poor stars. Recently, \cite{2018ApJ...857....2R} investigated the Cu abundances in six
late-type metal-poor stars using the UV \ion{Cu}{2} lines that were supposedly free of the NLTE influence. They showed that the Cu abundances
derived from \ion{Cu}{1} and \ion{Cu}{2} lines differed from each other by 0.36\,dex on average. Reanalyzing the Cu abundances of these six stars,
\cite{2018MNRAS.480..965K} showed that their NLTE consideration of the \ion{Cu}{1} lines removed the disparity between the abundances obtained from
\ion{Cu}{1} and \ion{Cu}{2} lines. 

Reviewing previous works, it is easy to conclude that observable \ion{Cu}{1} lines for low-metallicity stars
are susceptible to the influence of NLTE effects and an accurate and detailed Cu model atom needs to be defined for NLTE calculations.
Although both \cite{2014ApJ...782...80S} and \cite{2018MNRAS.473.3377A} built their respective Cu atomic models for NLTE calculations, the NLTE corrections
in these two models are slightly different: \cite{2018MNRAS.473.3377A} obtained higher NLTE corrections compared to those given by 
\cite{2018ApJ...862...71S}. The accuracy of NLTE calculations depends on the important inelastic collision processes with hydrogen.
For collisions with hydrogen, both studies used the so-called Drawin formula \citep{1968ZPhy..211..404D,1969ZPhy..225..483D,1984A&A...130..319S}. 
However, from comparisons with full quantum calculations for several elements, including Li, O, Na, Mg, Ca and Ni atoms
\citep{1999PhRvA..60.2151B,2010PhRvA..81c2706B,2012PhRvA..85c2704B,2018ApJ...867...87B,2019MNRAS.487.5097B,2003PhRvA..68f2703B,2011JPhB...44c5202G,2022ApJ...926..173V}, it is found that the classic formula is not reliable
and fails to describe the underlying mechanisms existing in inelastic collisions with hydrogen \citep{2011A&A...530A..94B}. A common approach 
to correcting for inadequacies in the Drawin recipe is to scale the rate coefficients by an empirical factor S$_H$. We note that a scale of
S$_H$\,=\,0.1 was used by \cite{2014ApJ...782...80S}, which was later tested in a series of works \citep{2015ApJ...802...36Y, 2016A&A...585A.102Y, 2018ApJ...862...71S, 2019ApJ...875..142X, 2020RAA....20..131X}.
Therefore, the differences in both copper atomic models may lead to the discrepancy between the two works.

As full quantum calculations are time-consuming, especially for the quantum-chemical data, the above approximate approaches are often in use. Until recently,
\cite{2021MNRAS.501.4968B} investigated the inelastic processes in low-energy Cu\,+\,H and Cu$^+$\,+\,H$^-$ collisions (306 partial processes in total) considering
the fine-structure effects. And it provides us with an opportunity to explore the influence on the NLTE effects of copper lines when
using accurate data.

The present study is motivated by the appearance of quantum-mechanical rate coefficients for inelastic collisions with hydrogen
performed by \cite{2021MNRAS.501.4968B}. In this paper, we investigate how the use of data from \cite{2021MNRAS.501.4968B}
affects the derived copper abundances for a sample of metal-poor stars with metallicity of $-$2.59\,dex $\le$ [Fe/H] $\le$ $-$0.95\,dex,
and on the trend of [Cu/Fe] versus [Fe/H]. Meantime, we aim to independently estimate the reliability of our old and 
newly updated copper atomic model. The observations are briefly summarized in Section \ref{sec:observations}.
Stellar parameters, atomic line data, model atmospheres, and atomic model are described in  Section \ref{sec:method}.
The results and discussion are given in Section \ref{sec:results}. And our conclusions are presented in Section \ref{sec:conclusion}.

\section{Observations} \label{sec:observations}
The stellar sample analyzed is the same as in \cite{2018ApJ...857....2R} and \cite{2018MNRAS.480..965K}, and was detailedly described in
the above papers. The UV spectra were downloaded from the Mikulski Archive for Space Telescopes (MAST), which were collected 
with the Space Telescope Imaging Spectrograph \citep[STIS;][]{1998ApJ...492L..83K,1998PASP..110.1183W} on board the Hubble Space Telescope (\emph{HST}). 
The resolved power was \emph{R} = 114\,000. Most of the investigated \ion{Cu}{1} lines and all of the \ion{Cu}{2} lines are 
situated in the UV spectra. For all program stars, the following lines are detectable in the spectral range: \ion{Cu}{1} 2024{\AA}
and \ion{Cu}{2}\,2037, 2054, 2104, 2112, 2126 and 2148{\AA}. Besides, for HD\,84937, HD\,94028 and HD\,140283, the following lines
also can be found in those spectra: \ion{Cu}{1} 2165, 2199, 2225, 2227, and 2230\,{\AA} and \ion{Cu}{2} 2189 and 2247\,{\AA}.

The high-resolution UVES optical spectra have been used to supplement the UV spectra and they were obtained from the European
Southern Observatory (ESO) Science Archive Facility. This enables us to employ two resonant UV lines of 3247 and 3274\,{\AA}. 
For HD\,76932 and HD\,94028, the optical subordinate lines of 5105, 5153 and 5218\,{\AA} are also included in our analysis.

\section{Method of calculation} \label{sec:method}
\subsection{Stellar Parameters} \label{subsec:parameters}  
The stellar parameters of our program stars are adopted from \cite{2018ApJ...857....2R}, and they have also been used by \cite{2018MNRAS.480..965K}.
Briefly, the effective temperature $T_\mathrm{eff}$ was calculated from the \cite{2010A&A...512A..54C} metallicity-dependent
color-$T_\mathrm{eff}$ calibrations. The surface gravity log\,\emph{g} was obtained from fundamental relations
 \citep[][see formula 1]{2018ApJ...857....2R}, which was based on the $\it{Hipparcos}$ and $\it{Tycho}$-2 $\it{Gaia}$ parallaxes
and the masses estimated by \cite{2010A&A...512A..54C}. It needs to be pointed out that the log\,\emph{g} values calculated from 
the newly released Gaia EDR3 data are very similar to what is used here, and the differences are less than 0.02\,dex. 
The metallicity [Fe/H] and microturbulence velocity $\xi$ were determined via an iterative process with \ion{Fe}{1} and \ion{Fe}{2} lines. 
For the reader's convenience, the stellar parameters are repeated in Table\,\ref{tab:stellar parameters}.

\begin{deluxetable}{lcccccccc}[!]
\tablenum{1}
\tablecaption{Stellar parameters for the studied stars  \label{tab:stellar parameters}}
\tablehead{
\colhead{Star} & \colhead{} & \colhead{$T_\mathrm{eff}$} & \colhead{} & \colhead{log\,\emph{g}} & \colhead{} & \colhead{[Fe/H]} & \colhead{} & 
\colhead{$\xi$}\\
\colhead{}  & \colhead{} & \colhead{(K)} & \colhead{} & \colhead{(cgs)} & \colhead{} & \colhead{(dex)} & \colhead{} & \colhead{(km $s^{-1}$)}
}
\startdata
HD\,19445 & & 6070 & & 4.44 & & $-$2.12 & &1.60 \\
HD\,76932 & & 5945 & & 4.17 & &$-$0.95 & &1.10\\
HD\,84937 & & 6427 & & 4.14 & &$-$2.16 & &1.45 \\
HD\,94028 & & 6097 & & 4.34 & &$-$1.52 & &1.30 \\
HD\,140283 & & 5766 & & 3.64 & &$-$2.59 & &1.30 \\
HD\,160617 & & 6050 & & 3.91 & &$-$1.89 & &1.50 \\
\enddata
\end{deluxetable}

\subsection{Atomic Line Data} \label{subsec:linedata}	 

The copper lines used in this study are taken from \cite{2018MNRAS.480..965K}, and the wavelength ($\lambda$), low excitation
potential ($E_\mathrm{low}$) and oscillator strengths (log\,\emph{gf}) are listed in Table\,\ref{tab:lines}. It needs to be
pointed out that we do not use all the lines, and the \ion{Cu}{1} 2214\,{\AA} line is discarded in our analysis because it's blended with other lines for our program stars. 
For the optical copper lines, the ratio between the two copper isotopes
($^{63}$Cu and $^{65}$Cu) is adopted as 0.69\,:\,0.31\citep{2009ARA&A..47..481A}, and the hyperfine structures have been taken into account. 
Table \ref{tab:lines} presents the mean wavelengths and average oscillator strengths. A detailed list of the wavelengths
and oscillator strengths can be found in \cite{2014ApJ...782...80S,2018ApJ...862...71S}. While for the UV lines, the hyperfine
structures have not been considered, and the oscillator strengths are adopted from 
\cite{2018MNRAS.480..965K}. As described in their work, the log\,\emph{gf} values of the \ion{Cu}{2} lines are taken from \cite{2005PhRvA..72a2507D},
and those for most neutral UV copper lines are taken from \cite{2011CaJPh..89..417K}. In addition, for the \ion{Cu}{1} 2024\,{\AA}
and 2165\,{\AA} lines, the data are respectively from \cite{1980PhyS...21...47L} and \cite{1991ApJS...77..119M}.

\begin{deluxetable}{lcclccrcc}[!]
\tablenum{2}
\tablecaption{Atomic Data of the copper lines  \label{tab:lines}}
\tablehead{
\colhead{$\lambda$(\AA)} & &  & \colhead{$E_\mathrm{low}$(ev)} &  & & \colhead{log\,\emph{gf}} &
}
\startdata
\ion{Cu}{1}\\
2024.325  & & & 0.0 & & &$-$1.75  \\
2024.338  & & & 0.0 & & &$-$1.46  \\
2165.096  & & & 1.3889 & & &$-$0.84  \\
2199.586  & & & 1.3889 & & &0.45 \\
2199.754  & & & 1.6422 & & &0.34  \\
2225.705  & & & 0.0 & & &$-$1.20  \\
2227.776  & & & 1.6422 & & &0.46  \\
2230.086  & & & 1.3889 & & &0.64  \\
\\
3247.54 & & &0.0 & & &$-$0.21  \\
3273.95 & & &0.0 & & &$-$0.50  \\
5105.54 & & &1.3890 & & &$-$1.64  \\
5153.23 & & &3.7859 & & &$-$0.01  \\
5218.20 & & &3.8167 & & &0.28   \\
\hline
\ion{Cu}{2}\\
2037.127 & & &2.8327 & & &$-$0.28  \\
2054.979 & & &2.8327 & & &$-$0.30  \\
2104.796 & & &2.9754 & & &$-$0.60  \\
2112.100 & & &3.2564 & & &$-$0.14  \\
2126.044 & & &2.8327 & & &$-$0.32  \\
2148.984 & & &2.7188 & & &$-$0.49  \\
2189.630 & & &3.2564 & & &$-$0.39  \\
2247.003 & & &2.7188 & & &0.10  \\
\enddata
\end{deluxetable}

\begin{figure*}[!]
\centering
\includegraphics[width=0.99\textwidth,angle=0]{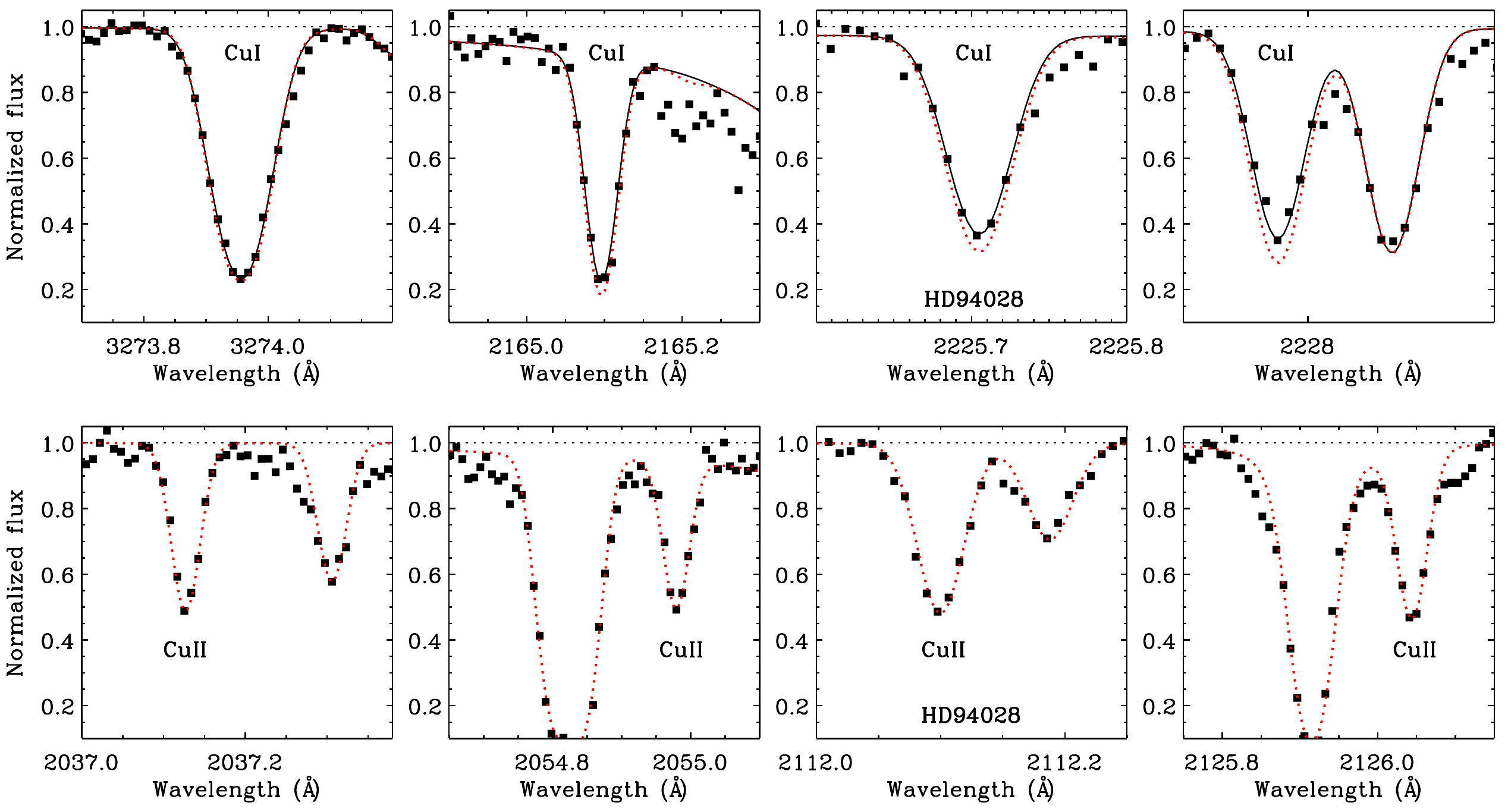}
\caption{Synthetic profiles of \ion{Cu}{1} lines (upper panel) and \ion{Cu}{2} lines (lower panel) for HD\,94028. The filled
squares are the observed spectra, the black solid line is the best-fitting synthesis (NLTE synthesis), and the red dotted line
is the LTE profile with the same [Cu/Fe] relative to the NLTE.
\label{fig.profiles}\\}
\end{figure*}

\subsection{Model atmospheres} \label{subsec:atmosphere}

For each star, we compute the 1D LTE MARCS atmospheric models, which are obtained by interpolation in the grid of
plane-parallel MARCS models for 3.5\,dex\,$\leqslant$\,log\,\emph{g}\,$\leqslant$\,5.5\,dex \citep[see][for details]{2008A&A...486..951G}. 
For each individual model, the $\alpha$-enhancement is under consideration, as described below:

$[\alpha/\text{Fe}]=\left\{
   \begin{array}{lcl}
       \text{0.0} & \text{if} & \hspace{0.3cm}0.0 < \text{[Fe/H]} \\
       \text{0.4}$$\times$$|\text{[Fe/H]}| & \text{if} & -1.0 \leqslant \text{[Fe/H]}
       \leqslant \hspace{0.28cm}0.0 \\
       \text{0.4} & \text{if} & \hspace{1.2cm}\text{[Fe/H]} < -1.0
   \end{array}
   \right.
 $

\subsection{Updated Atomic Model} \label{subsec:atomic_model}

For our NLTE calculations, we adopt the copper atomic model from \cite{2014ApJ...782...80S} as a basis. It contains 97 energy levels
(96 \ion{Cu}{1} levels and the \ion{Cu}{2} ground state) and 1089 line transitions of \ion{Cu}{1}. The fine structures 
of the levels with low excitation energy are included, and the transition oscillator strengths and the photoionization cross sections from\,\cite{2014ApJS..211...30L}
are introduced into our atomic model. To calculate the excitation of allowed and forbidden transitions by electron collisions, we use the 
formulae of \cite{1962ApJ...136..906V} and \cite{Allen1973}, respectively. And the formula of \cite{Seaton1962} is used to calculate the
collisional ionization rates. 

A novelty of this study is that we take into account the mutual neutralization, ion-pair formation, excitation, and de-excitation 
inelastic processes in copper-hydrogen collisions according to the data from \cite{2021MNRAS.501.4968B}  (hereafter the new model). Previously, 
for the calculation of the inelastic collisions with hydrogen, we used the Drawin's formula \citep{1968ZPhy..211..404D,1969ZPhy..225..483D}, which 
was described by \cite{1984A&A...130..319S} with a scaling factor of S$_H$=0.1 following the suggestion of \cite{2014ApJ...782...80S} (hereafter the old model).

In order to solve the statistical equilibrium and the coupled radiative transfer equations, we employ the revised DETAIL code \citep{BG1985},
which is based on an accelerated lambda iteration method \citep{1991A&A...245..171R,1992A&A...262..209R}, to calculate the NLTE occupation numbers for the copper atomic model.
Then, the obtained departure coefficients are used to compute the synthetic line profiles via the SIU program \citep{Reetz1991}.

\section{Results and Discussion} \label{sec:results}
\subsection{Stellar Copper Abundances} \label{subsec:abundances}

The LTE and NLTE copper abundances in our six program stars are derived through the spectral synthesis method based on the \ion{Cu}{1} and \ion{Cu}{2} lines. 
Abundances determination in this way is an iterative process and we change the copper abundances until the differences
between the synthetic spectra and the observed ones are minimized.

Figure\,\ref{fig.profiles} illustrates the observed and synthetic line profiles of eight copper lines: the upper panel
shows the profiles of four \ion{Cu}{1} lines, and the lower panel presents the four \ion{Cu}{2} line profiles of HD\,94028.
In this figure, the LTE and NLTE profiles of the copper lines are indicated by the red dotted lines and the 
black solid lines, respectively. It needs to be pointed out that the LTE and NLTE profiles are synthesized with the same abundance for the \ion{Cu}{1} line
in the upper panel of Figure\,\ref{fig.profiles}, and the LTE assumptions underestimate the Cu abundances.

\begin{deluxetable*}{lccccccccccccccc}[!]
\tablenum{3}
\tablecaption{[Cu/Fe] for \ion{Cu}{1} lines based on new and old copper atomic model  \label{tab:CuI abundances}}
\tablehead{
\colhead{Star}  &\colhead{3247{\AA}} & \colhead{3274{\AA}} & \colhead{5105{\AA}} & \colhead{5153{\AA}} & \colhead{5218{\AA}} & \colhead{2024{\AA}} & \colhead{2165{\AA}} & \colhead{2199{\AA}} & \colhead{2225{\AA}} & \colhead{2227{\AA}} & \colhead{2230{\AA}} & \colhead{Mean+$\sigma$} & \colhead{$\Delta_\mathrm{NLTE}$}
}
\startdata
HD\,19445 & $-$0.68 & $-$0.73 & $\cdots$ & $\cdots$ & $\cdots$ & $\cdots$ & $\cdots$ & $\cdots$ & $\cdots$ & $\cdots$ & $\cdots$ & $-$0.71$\pm$0.02 & \\
          & $-$0.40 & $-$0.41 & $\cdots$ & $\cdots$ & $\cdots$ & $\cdots$ & $\cdots$ & $\cdots$ & $\cdots$ & $\cdots$ & $\cdots$ & $-$0.41$\pm$0.01 & 0.30\\
          & $-$0.31 & $-$0.33 & $\cdots$ & $\cdots$ & $\cdots$ & $\cdots$ & $\cdots$ & $\cdots$ & $\cdots$ & $\cdots$ & $\cdots$ & $-$0.32$\pm$0.01 & 0.39\\
HD\,76932 & $-$0.20 & $-$0.18 & $-$0.21 & $-$0.22 & $-$0.27 & $-$0.20 & $\cdots$ & $\cdots$ & $\cdots$ & $\cdots$ & $\cdots$ & $-$0.21$\pm$0.03 &\\
          & $-$0.18 & $-$0.17 & $-$0.11 & $-$0.16 & $-$0.22 & $-$0.14 & $\cdots$ & $\cdots$ & $\cdots$ & $\cdots$ & $\cdots$ & $-$0.16$\pm$0.04 & 0.05\\
          & $-$0.17 & $-$0.16 & $-$0.06 & $-$0.12 & $-$0.17 & $-$0.10 & $\cdots$ & $\cdots$ & $\cdots$ & $\cdots$ & $\cdots$ & $-$0.13$\pm$0.04 & 0.08\\
HD\,84937 & $-$0.78 & $-$0.74 & $\cdots$ & $\cdots$ & $\cdots$ & $\cdots$ & $-$0.72 & $-$0.74 & $-$0.70 & $-$0.72 & $-$0.75 & $-$0.74$\pm$0.03 &\\
          & $-$0.35 & $-$0.32 & $\cdots$ & $\cdots$ & $\cdots$ & $\cdots$ & $-$0.35 & $-$0.38 & $-$0.36 & $-$0.36 & $-$0.35 & $-$0.35$\pm$0.02 & 0.39\\
          & $-$0.30 & $-$0.28 & $\cdots$ & $\cdots$ & $\cdots$ & $\cdots$ & $-$0.31 & $-$0.35 & $-$0.33 & $-$0.33 & $-$0.32 & $-$0.32$\pm$0.02 & 0.42\\
HD\,94028 & $-$0.45 & $-$0.47 & $-$0.50 & $\cdots$ & $\cdots$ & $-$0.45 & $-$0.46 & $-$0.49 & $-$0.45 & $-$0.50 & $-$0.49 & $-$0.47$\pm$0.02 &\\
          & $-$0.37 & $-$0.39 & $-$0.28 & $\cdots$ & $\cdots$ & $-$0.29 & $-$0.27 & $-$0.33 & $-$0.30 & $-$0.31 & $-$0.28 & $-$0.31$\pm$0.04 & 0.16\\
          & $-$0.30 & $-$0.33 & $-$0.23 & $\cdots$ & $\cdots$ & $-$0.23 & $-$0.21 & $-$0.29 & $-$0.24 & $-$0.26 & $-$0.23 & $-$0.26$\pm$0.04 & 0.21\\
HD\,140283 & $-$0.71 & $-$0.56 & $\cdots$ & $\cdots$ & $\cdots$ & $\cdots$ & $-$0.53 & $-$0.54 & $-$0.54 & $-$0.48 & $-$0.58 & $-$0.56$\pm$0.07 &\\
           & $-$0.32 & $-$0.18 & $\cdots$ & $\cdots$ & $\cdots$ & $\cdots$ & $-$0.19 & $-$0.24 & $-$0.21 & $-$0.17 & $-$0.27 & $-$0.23$\pm$0.05 &0.33\\
           & $-$0.24 & $-$0.11 & $\cdots$ & $\cdots$ & $\cdots$ & $\cdots$ & $-$0.12 & $-$0.19 & $-$0.15 & $-$0.12 & $-$0.21 & $-$0.16$\pm$0.05 &0.40\\
HD\,160617 & $-$0.69 & $-$0.62 & $\cdots$ & $\cdots$ & $\cdots$ & $\cdots$ & $\cdots$ & $\cdots$ & $\cdots$ & $\cdots$ & $\cdots$ & $-$0.66$\pm$0.04 &\\
           & $-$0.40 & $-$0.33 & $\cdots$ & $\cdots$ & $\cdots$ & $\cdots$ & $\cdots$ & $\cdots$ & $\cdots$ & $\cdots$ & $\cdots$ & $-$0.37$\pm$0.04 &0.29\\
           & $-$0.33 & $-$0.27 & $\cdots$ & $\cdots$ & $\cdots$ & $\cdots$ & $\cdots$ & $\cdots$ & $\cdots$ & $\cdots$ & $\cdots$ & $-$0.30$\pm$0.03 &0.36\\
\enddata
\tablecomments{The LTE abundances of each star are shown in the first row, while the NLTE results respectively based on new and old copper atomic model are indicated in the second and third row. }
\end{deluxetable*}

\begin{figure}[htb!]
\centering
\includegraphics[width=0.51\textwidth,angle=0]{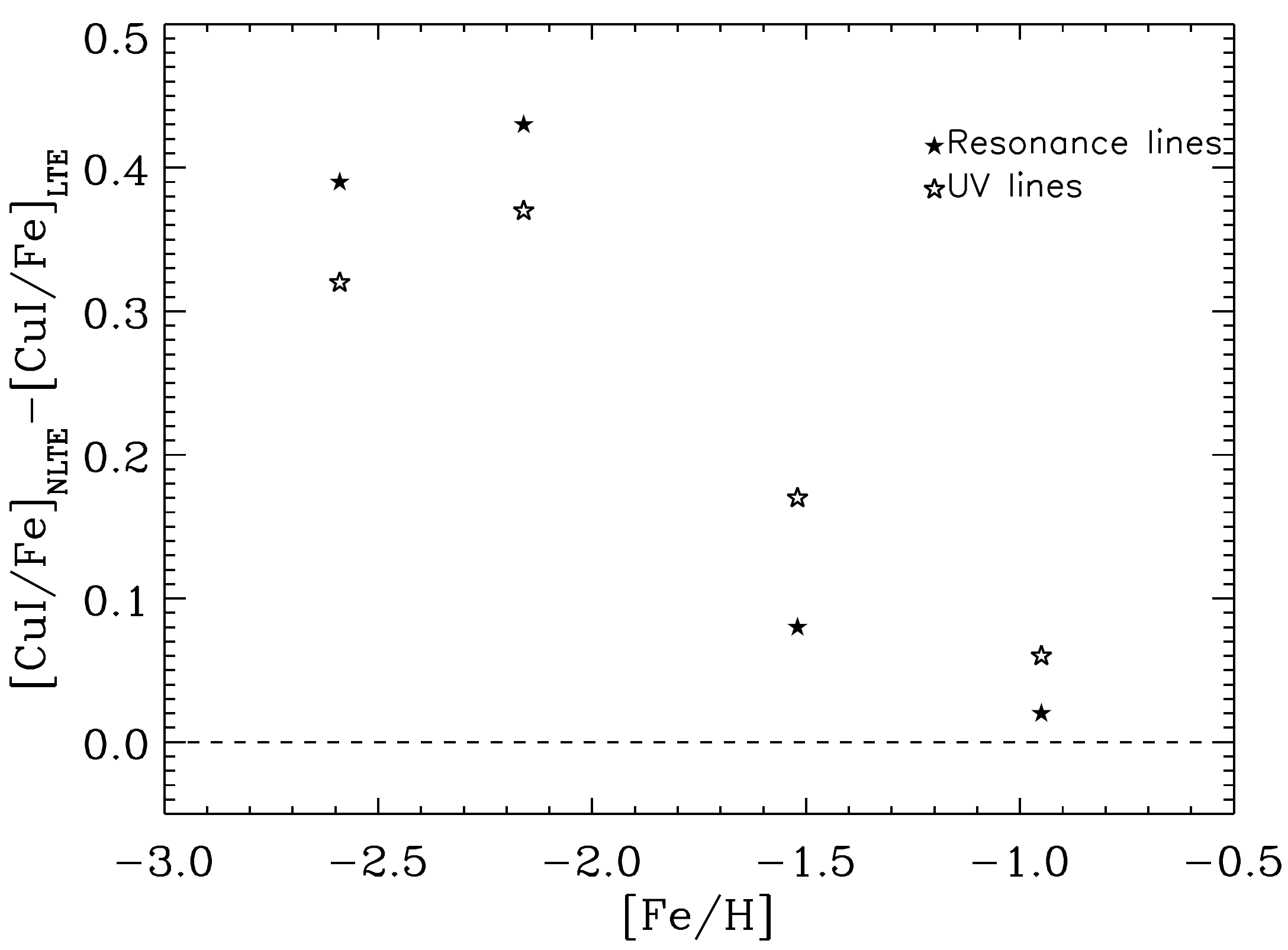}
\includegraphics[width=0.37\textwidth,angle=90]{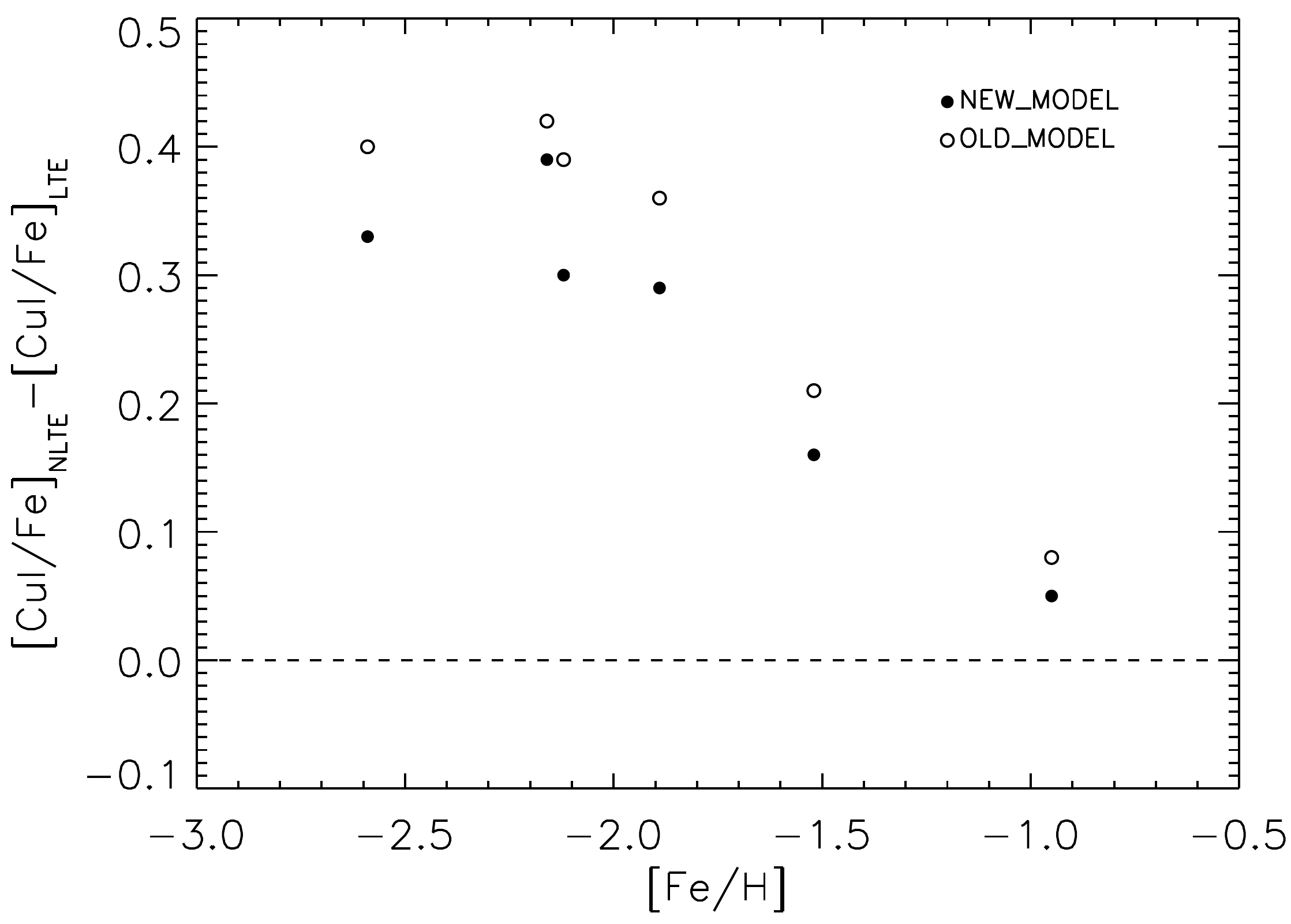}
\caption{The NLTE corrections to the LTE copper abundances derived from \ion{Cu}{1} lines as a function of [Fe/H].
\label{fig.nltecorrection}}
\end{figure}

In Table\,\ref{tab:CuI abundances}, we present the [Cu/Fe] values derived from the line profile fitting of
individual \ion{Cu}{1} lines and the mean results along with the standard deviation. The LTE abundances 
of each star are shown in the first row, while the NLTE results based on the new and old copper 
atomic model are indicated in the second and third row, respectively. Considering that the copper abundances are mostly derived based on 
the UV copper lines, the uncertainty in abundance determination mainly comes from the indeterminacy of continuum determination,
which can vary from 0.05 to 0.1\,dex for a 1\% shift from the true continuum flux level.
Through inspection of Table\,\ref{tab:CuI abundances}, it can be found that the NLTE corrections,
$\Delta_\mathrm{NLTE}$ = [\ion{Cu}{1}/Fe]$_\mathrm{NLTE}$ - [\ion{Cu}{1}/Fe]$_\mathrm{LTE}$, are evident 
for both copper atomic models, and the effects differ from line to line even for the same star. The NLTE corrections vary from 0.05 to 0.39\,dex and 0.08 to 0.42\,dex,
 for the new and old copper model, respectively.
In Figure \ref{fig.nltecorrection}, the upper panel presents the NLTE corrections for resonance (filled pentagrams)
and UV \ion{Cu}{1} lines (open pentagrams) based on the new copper atomic model. It can be seen that the NLTE corrections decrease with increasing metallicity, which has
been reported in previous studies \citep[][]{2015ApJ...802...36Y,2018MNRAS.473.3377A,2018MNRAS.480..965K,2018ApJ...862...71S,2019ApJ...875..142X,2020RAA....20..131X}.
The NLTE corrections have no relation to the adopted \ion{Cu}{1} lines. The lower panel shows the NLTE corrections for our two copper atomic models (filled circles denote the new model, while open circles represent the old one), and it indicates that the same feature (i.e. the NLTE corrections decrease with increasing metallicity) exists. Theoretically, if the [Fe/H] increases, UV line blocking will increase rapidly, and the overionization
will become less important. Therefore, the NLTE correction, as expected, will decrease with increasing [Fe/H].
Meantime, it is clear that the NLTE corrections for the new model are lower than the old one,
indicating that the application of accurate atomic data leads to a decrease in the departure from LTE.

\subsection{Influence of Inelastic Collisions with Hydrogen on the Cu Abundance}

\begin{figure}[!]
\centering
\subfigure[]{
\label{fig.sub.1}
\includegraphics[width=0.35\textwidth,angle=90]{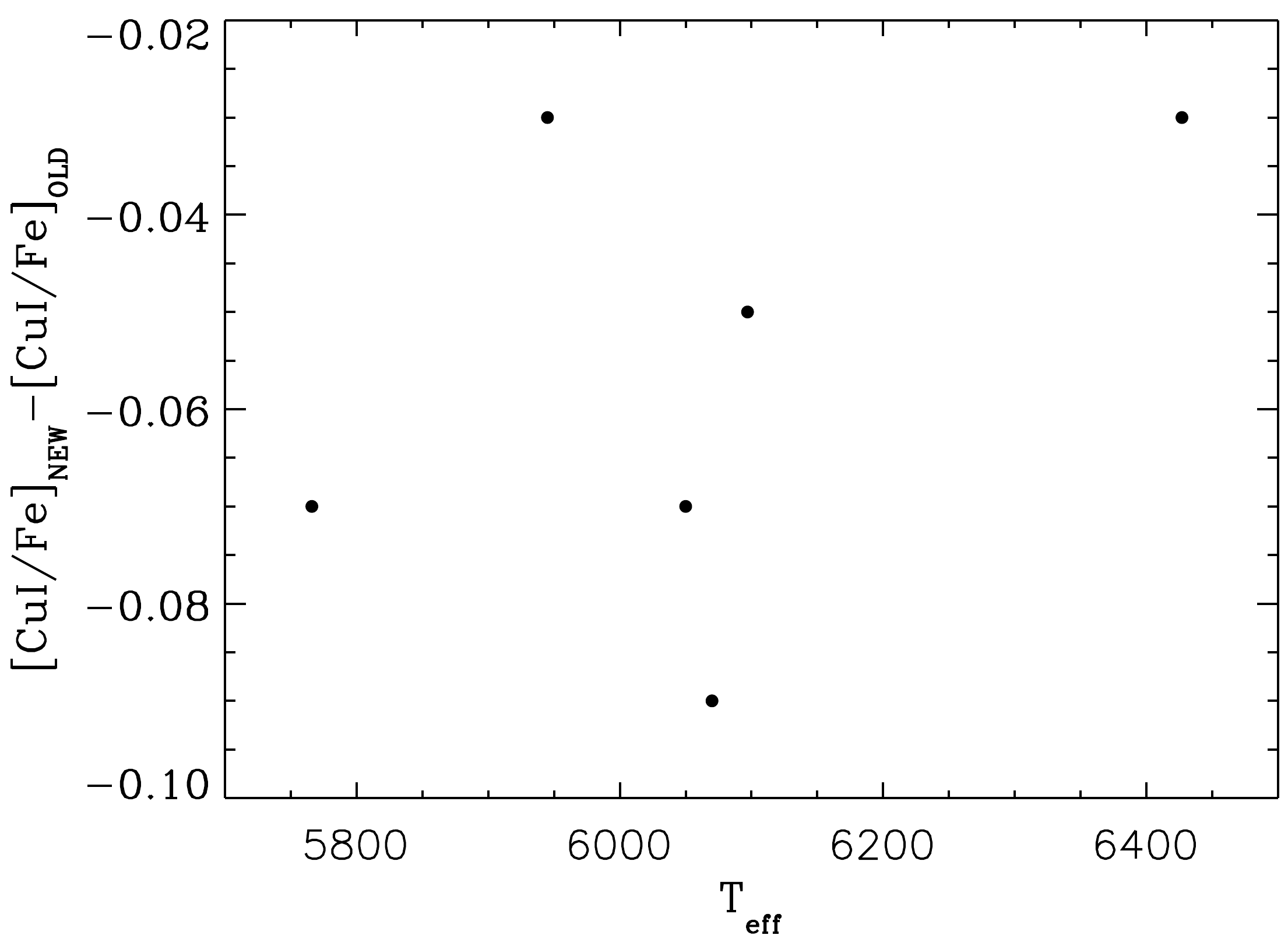}}
\subfigure[]{
\label{fig.sub.2}
\includegraphics[width=0.35\textwidth,angle=90]{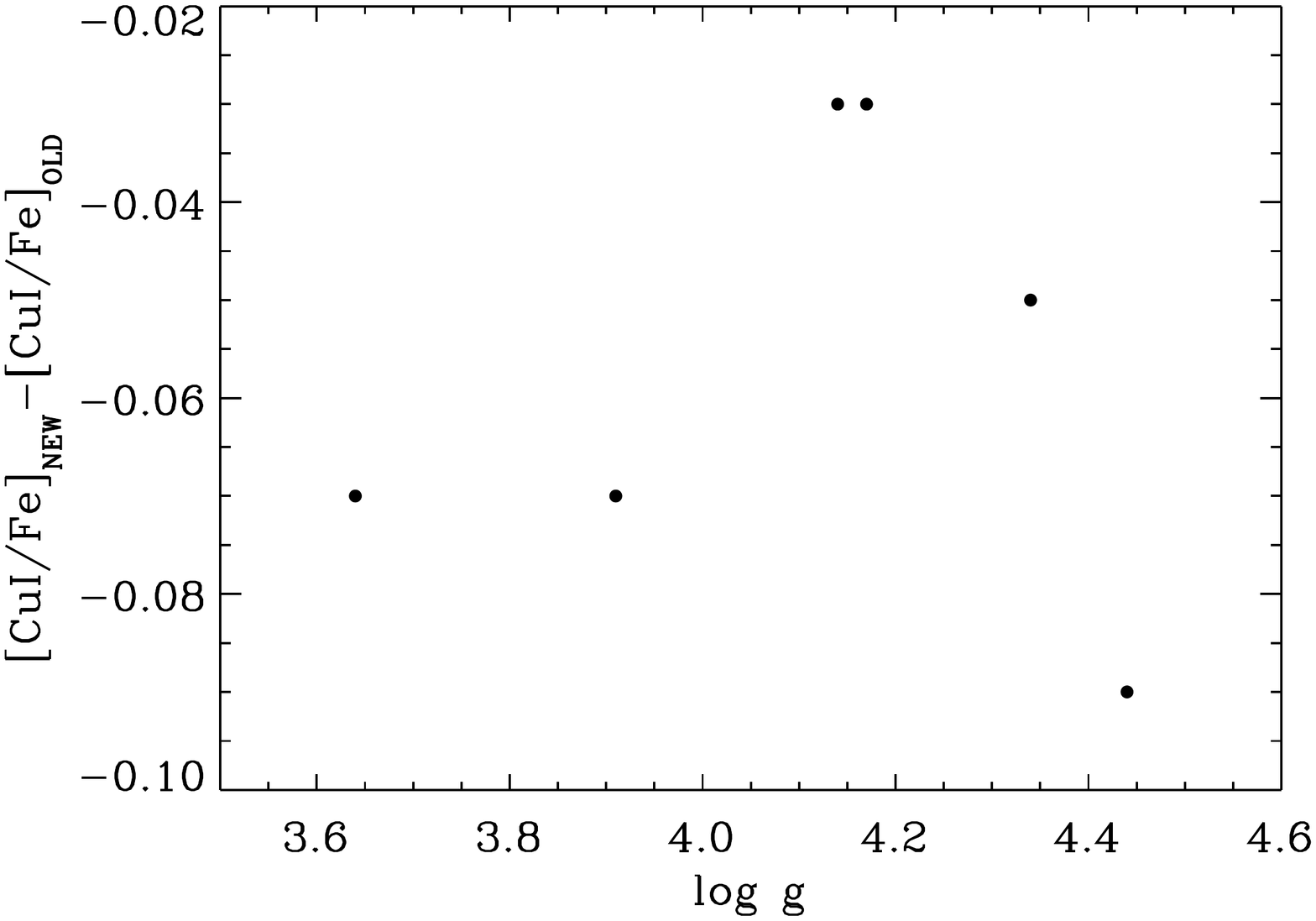}}
\subfigure[]{
\label{fig.sub.3}
\includegraphics[width=0.35\textwidth,angle=90]{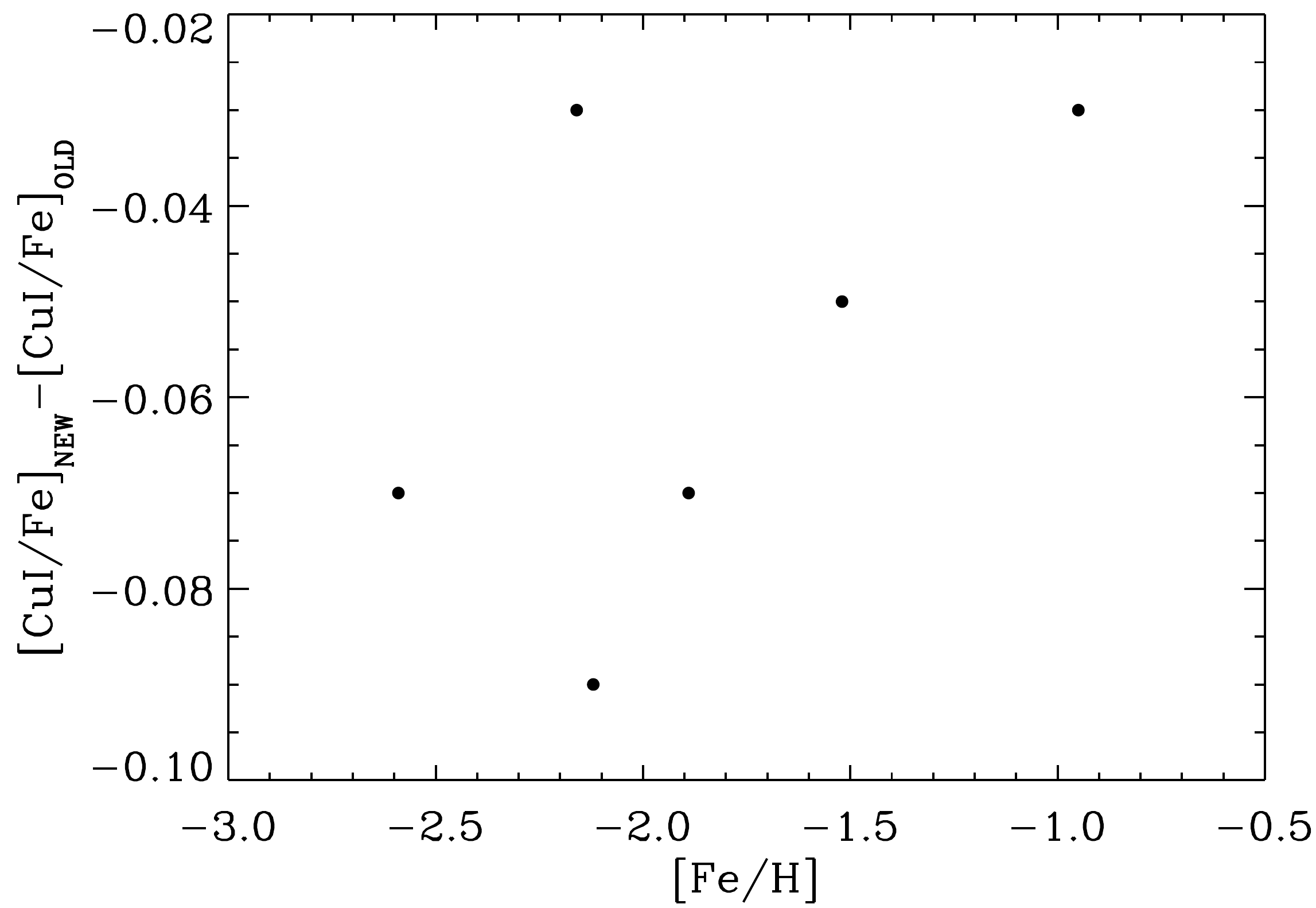}}
\caption{The mean differences between the NLTE abundances derived from the \ion{Cu}{1} lines for the new and old copper
atomic models as functions of the stellar atmospheric parameters.
\label{fig.parameters}}
\end{figure}

Figure\,\ref{fig.parameters} displays the mean differences between the NLTE Cu abundances derived from the \ion{Cu}{1}
lines for the new and old copper atomic models as functions of effective temperature, surface gravity and metallicity.

We find the difference in NLTE abundance depends on the metallicity with a range from $-$0.09 to $-$0.03\,dex. The difference between the NLTE abundances
derived with the new (accurate data) and old (approximate data) model diminishes with increasing metallicity, because, in more metal-rich stars, the collision with 
electron become more important.
Moreover, it can be seen that the differences are negative, which indicates that the application of accurate data for our program stars
leads to lower NLTE Cu abundances.

\subsection{Verify the Adopted Copper Atomic Model} \label{subsec:verify}

In Table\,\ref{tab:CuII abundances}, we give the LTE copper abundances determined with individual \ion{Cu}{2} lines.
The penultimate column shows the difference between the copper abundances derived from \ion{Cu}{1} (NLTE) and \ion{Cu}{2} (LTE)
lines for the new copper atomic model. The difference varies from $-$0.02 to 0.04\,dex, which can be explained by the 
standard deviation in the mean results. Theoretically, as the copper atoms are mainly remaining in the ionization state
in the metal-poor stars, the \ion{Cu}{2} lines are supposedly free of the NLTE influence. The nearly same abundance
both from the lines of two ionization stages suggests that our new atomic model is valid.

\begin{figure}[htb!]
\centering
\includegraphics[width=0.53\textwidth,angle=0]{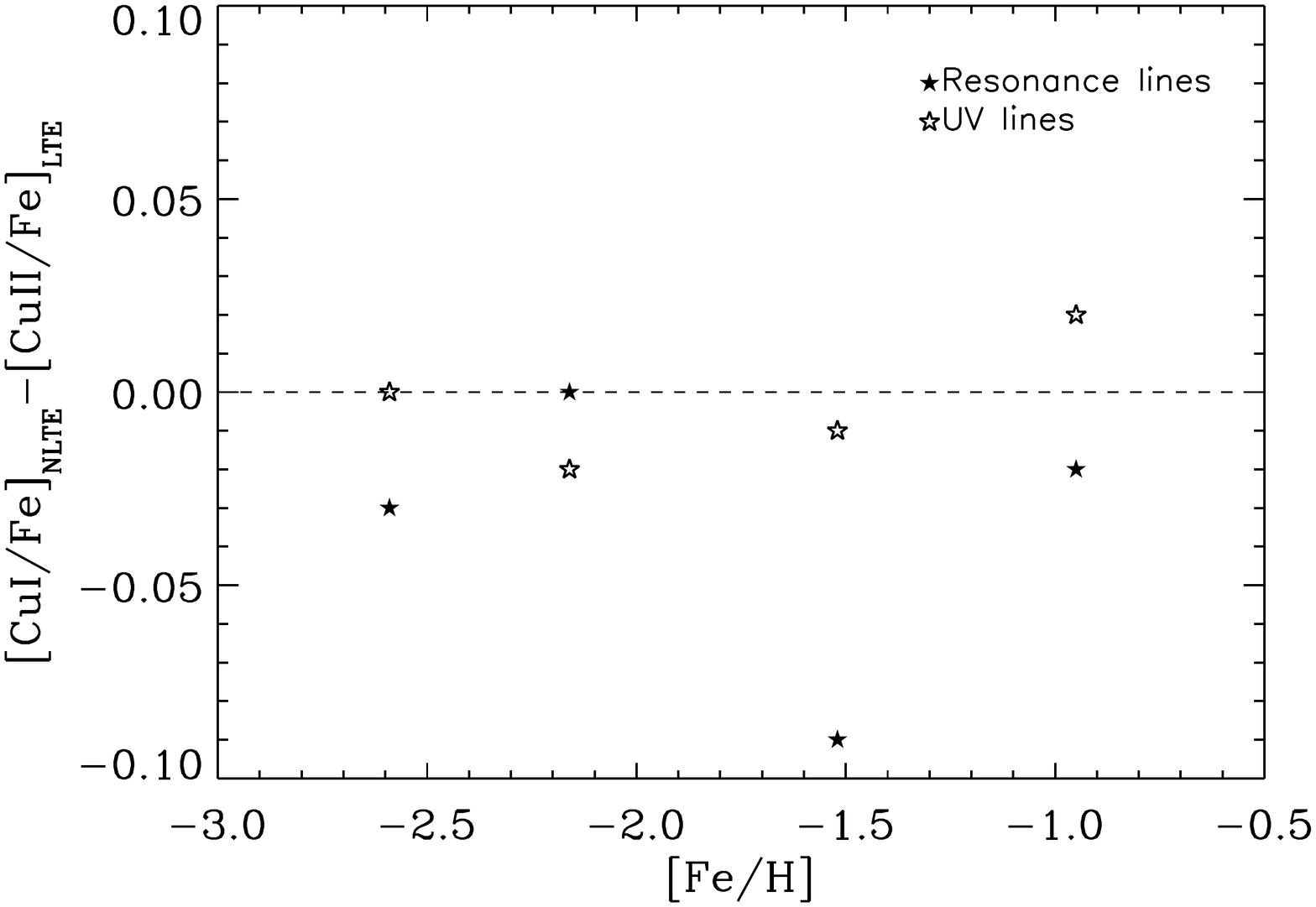}
\includegraphics[width=0.37\textwidth,angle=90]{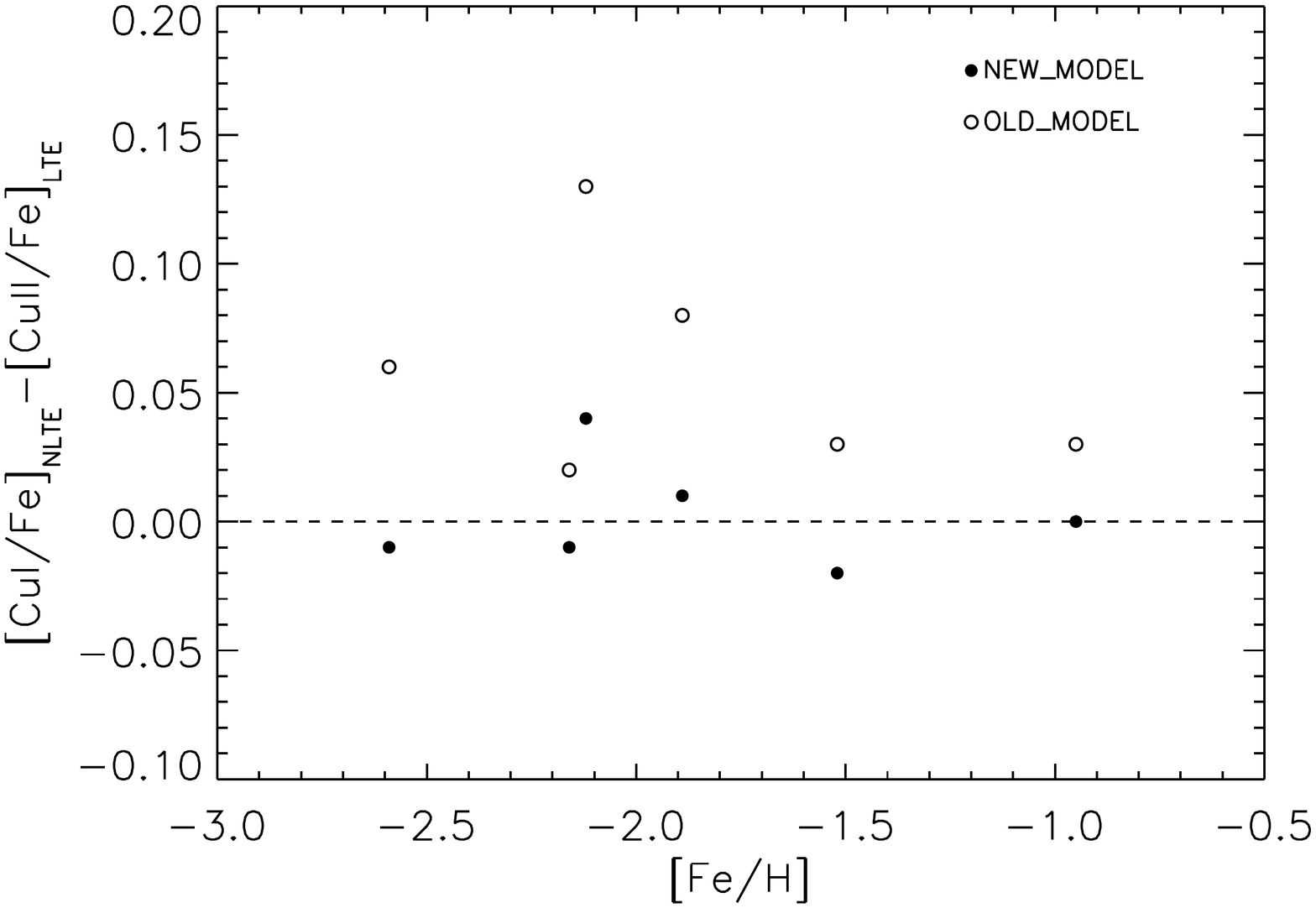}
\caption{Differences between the Cu abundances derived from the \ion{Cu}{1} (NLTE) and \ion{Cu}{2} (LTE) lines for our new and 
old copper atomic model. The same symbols are used as in Figure\,\ref{fig.nltecorrection}.
\label{fig.new_o_difference}\\}
\end{figure}

To more visually verify our adopted copper atomic model, we present the abundance differences as a function of [Fe/H] in
Figure\,\ref{fig.new_o_difference}. The upper panel shows the differences for the resonance and UV lines based on the new model.
Obviously, not only for the resonance lines but also for the UV lines, the same conclusion can be drawn as described in the above paragraph.
The lower panel presents the differences for our new and old models. It can be seen that the difference is negligible for both models
when the errors of abundances are considered. Besides, it is noted that although the inelastic collisions with hydrogen are obtained
with approximate formulas for the old model, our old atomic model is also valid to analyze the NLTE copper abundances to some extent.

\begin{deluxetable*}{lccccccccccccc}[!]
\tablenum{4}
\tablecaption{[Cu/Fe] for \ion{Cu}{2} under LTE conditions  \label{tab:CuII abundances}}
\tablehead{
\colhead{Star} & \colhead{2037{\AA}} & \colhead{2054{\AA}} & \colhead{2104{\AA}} & \colhead{2112{\AA}} & \colhead{2126{\AA}} & \colhead{2148{\AA}} & \colhead{2189{\AA}} & \colhead{2247{\AA}} & \colhead{mean+$\sigma$} & \colhead{[\ion{Cu}{1}/Fe]$-$[\ion{Cu}{2}/Fe]} & \colhead{[Cu/Fe]} &
}
\startdata
HD\,19445 & $-$0.50 & $-$0.48 & $-$0.42 & $-$0.42 & $-$0.49 & $-$0.40 & $\cdots$ & $\cdots$ & $-$0.45$\pm$0.04 & $+$0.04 & $-$0.44 \\
HD\,76932 & $-$0.21 & $-$0.16 & $-$0.13 & $\cdots$ & $-$0.15 & $\cdots$ & $\cdots$ & $\cdots$ & $-$0.16$\pm$0.03 & 0.00 & $-$0.16\\
HD\,84937 & $-$0.31 & $-$0.34 & $-$0.30 & $\cdots$ & $-$0.38 & $\cdots$ & $\cdots$ & $-$0.38 & $-$0.34$\pm$0.04 & $-$0.01 & $-$0.35 \\
HD\,94028 & $-$0.42 & $-$0.40 & $-$0.30 & $-$0.15 & $-$0.29 & $-$0.15 & $-$0.29 & $\cdots$ & $-$0.29$\pm$0.10 & $-$0.02 & $-$0.30 \\
HD\,140283 & $\cdots$ & $-$0.20 & $\cdots$ & $-$0.27 & $-$0.23 & $-$0.19 & $\cdots$ & $-$0.22 & $-$0.22$\pm$0.03 & $-$0.01 & $-$0.22 \\
HD\,160617 & $-$0.42 & $-$0.38 & $\cdots$ & $-$0.31 & $-$0.39 & $\cdots$ & $\cdots$ & $\cdots$ & $-$0.38$\pm$0.05 & $+$0.01 & $-$0.37 \\
\enddata
\end{deluxetable*}

\subsection{Comparison with Other Work}  \label{subsec:comparison}

Several groups have derived copper abundances for metal-poor stars under both LTE and NLTE assumptions. We compare our results
with those from other works and discuss the reasons for the differences.

\cite{2015ApJ...802...36Y, 2016A&A...585A.102Y}. They derived the copper abundances of a large number of metal-poor stars with 
the \ion{Cu}{1} lines at 5105.5, 5218.2, and 5782.1\,\AA. Their results are in good agreement with ours, and the average
difference between our NLTE abundances and theirs is $-$0.02 $\pm$ 0.03 for the two common stars.

\cite{2018MNRAS.473.3377A}. The authors developed a model atom for Cu and investigated the NLTE effects of \ion{Cu}{1} lines in 
very metal-poor stars. We have three objects in common with theirs, i.e., HD\,84937, \,HD94028, and HD\,140283. For HD\,84937,
their NLTE result is 0.15 dex higher than ours, while it is 0.13 for HD\,94028. For HD\,140283, both results agree well (0.03 dex). 
Since similar stellar parameters have been adopted in these two works, which can not explain the differences.
Besides, we note that both studies used different log\,\emph{gf} values for the five common optical lines, and their values are generally higher than ours
of 0.11 $\pm$ 0.07\,dex. Theoretically, it will lead to lower [Cu/Fe] ratios compared to ours. However,  
they derived higher NLTE copper abundances. The reason is that the different copper atomic model adopted in NLTE calculations for both works, e.g.,
for the inelastic collisions with hydrogen, \cite{2018MNRAS.473.3377A} used the Drawin formula, while our work uses the data from \cite{2021MNRAS.501.4968B}.

\cite{2018ApJ...857....2R}. They made a new test of copper abundances in late-type stars using ultraviolet \ion{Cu}{2} lines
and concluded that LTE underestimated the Cu abundances obtained from \ion{Cu}{1} lines. As their work has no NLTE results, we compare their
\ion{Cu}{2} LTE abundances with ours. For HD\,19445, HD\,76932, and HD\,94028, the results of both works agree well. While for HD\,160617, 
our LTE result is 0.16 dex higher than theirs, which can be explained by the difference of continuum determination. 
However, for HD\,84937 and HD\,140283, our results are 0.30 and 0.49 dex higher, respectively. We note that the numbers of copper lines used by both
works are different. It is noted that, except for the eight common copper lines from \cite{2018ApJ...857....2R}, we include another eleven UV lines.
For the eight common lines, \cite{2018ApJ...857....2R} adopted their log\,\emph{gf} values from the NIST Atomic Spectra Database, which are 
higher than ours. While our log\,\emph{gf} values are mainly adopted from previous literature (see Section\,3.2),
and have been tested by \cite{2018MNRAS.480..965K}. The average difference of the log\,\emph{gf} values is 0.08\,$\pm$\,0.06\,dex for the eight common lines. 
Thus, the difference between the log\,\emph{gf} values can lead to a discrepancy of about
0.1\,dex in copper abundance. Moreover, the statistical uncertainties include contributions from the uncertainty of line fitting 
and log\,\emph{gf} values have been listed in
Table\,7 of \cite{2018ApJ...857....2R}, and the values are 0.19 and 0.29\,dex. Considering the above mentioned factors, the large differences
can be explained.

\cite{2018ApJ...862...71S}. Based on the high-resolution, high signal-to-noise ratio UVES spectra, \cite{2018ApJ...862...71S} determined the copper abundances
of 29 metal-poor stars. There are four common stars between both studies. For HD\,76932 and HD\,84937, both results agree well, while for HD\,140283 and
HD\,160617, their abundances are 0.41 and 0.33\,dex lower than ours. For these two stars, we note that the adopted stellar parameters are different in both works,
which will lead to the large differences.
For HD\,140283, both studies used similar $T_\mathrm{eff}$ and log\,\emph{g} values, while their [Fe/H] and $\xi$ values are 0.18 and 0.2\,dex higher than ours.
For HD\,160617, although both works used the same $\xi$ (1.5\,km $s^{-1}$), their $T_\mathrm{eff}$ and log\,\emph{g} are 110\,K and 0.11\,dex lower than our values, 
and their [Fe/H] ratio is 0.11\,dex higher than ours. 
When we used the same parameters as \cite{2018ApJ...862...71S} to derive the copper abundances for the two stars, the differences will decrease to 0.13 and 0.10\,dex, which can be explained by the impact of continuum determination.

\cite{2018MNRAS.480..965K}. Using the same stars and stellar parameters with \cite{2018ApJ...857....2R}, \cite{2018MNRAS.480..965K} checked the consistency
between the copper abundance for ultraviolet \ion{Cu}{2} lines and \ion{Cu}{1} lines, and showed that their NLTE results of the \ion{Cu}{1}
lines removed disagreement between these two sets of lines presented in the former work. For the six common stars, the average difference between
their [Cu\,I/Fe] values and ours is 0.14 $\pm$ 0.07\,dex, while it is 0.05 $\pm$ 0.11\,dex between their [Cu\,II/Fe] values and our results. 
As we note they adopted the same copper atomic model with \cite{2018MNRAS.473.3377A}, and their [Cu\,I/Fe] ratios in NLTE assumptions are
systematically higher than ours, which is due to the difference between adopted copper atomic models.

\begin{figure*}[htb!]
\centering
\includegraphics[width=0.49\textwidth,angle=0]{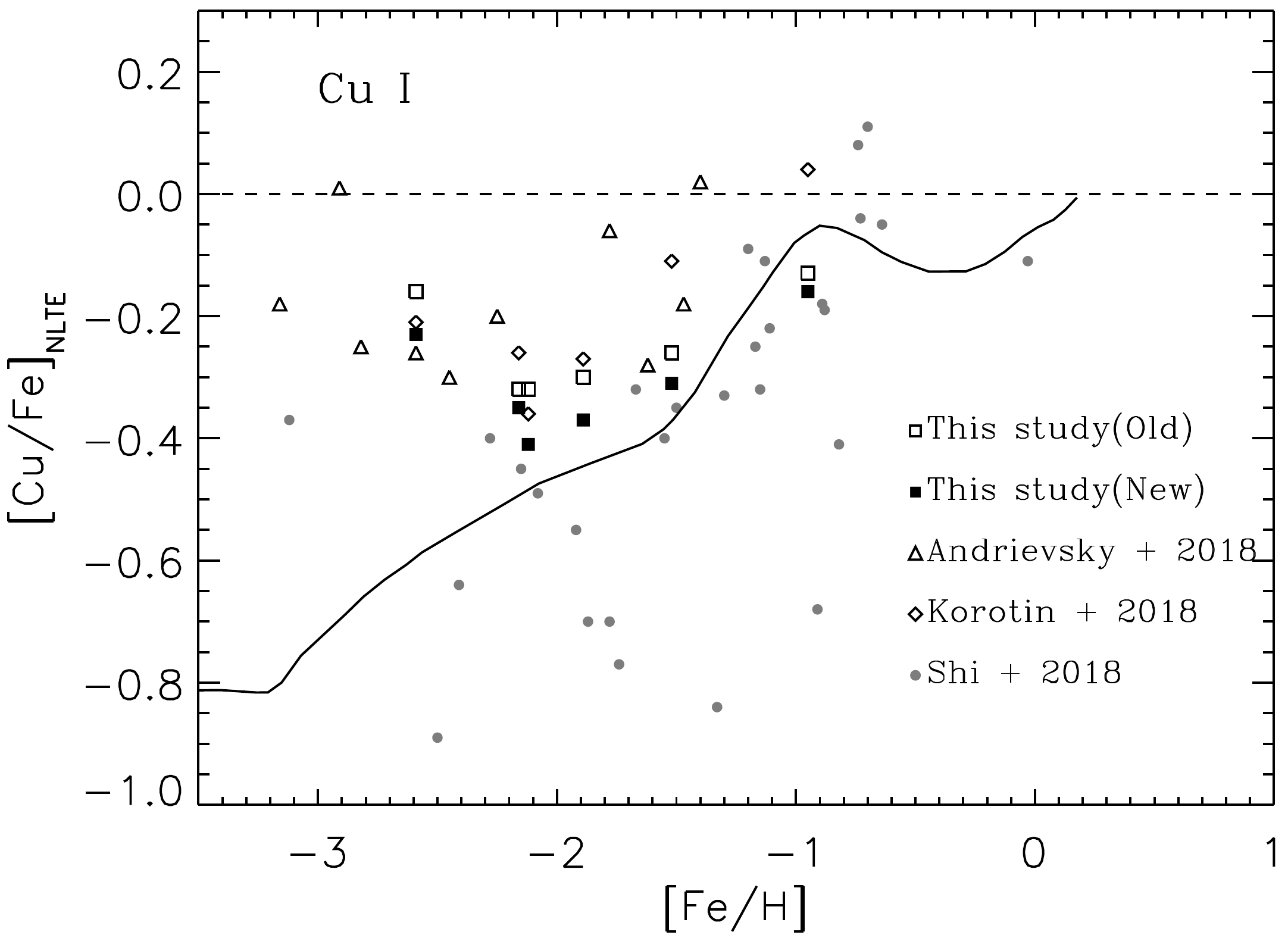}
\includegraphics[width=0.49\textwidth,angle=0]{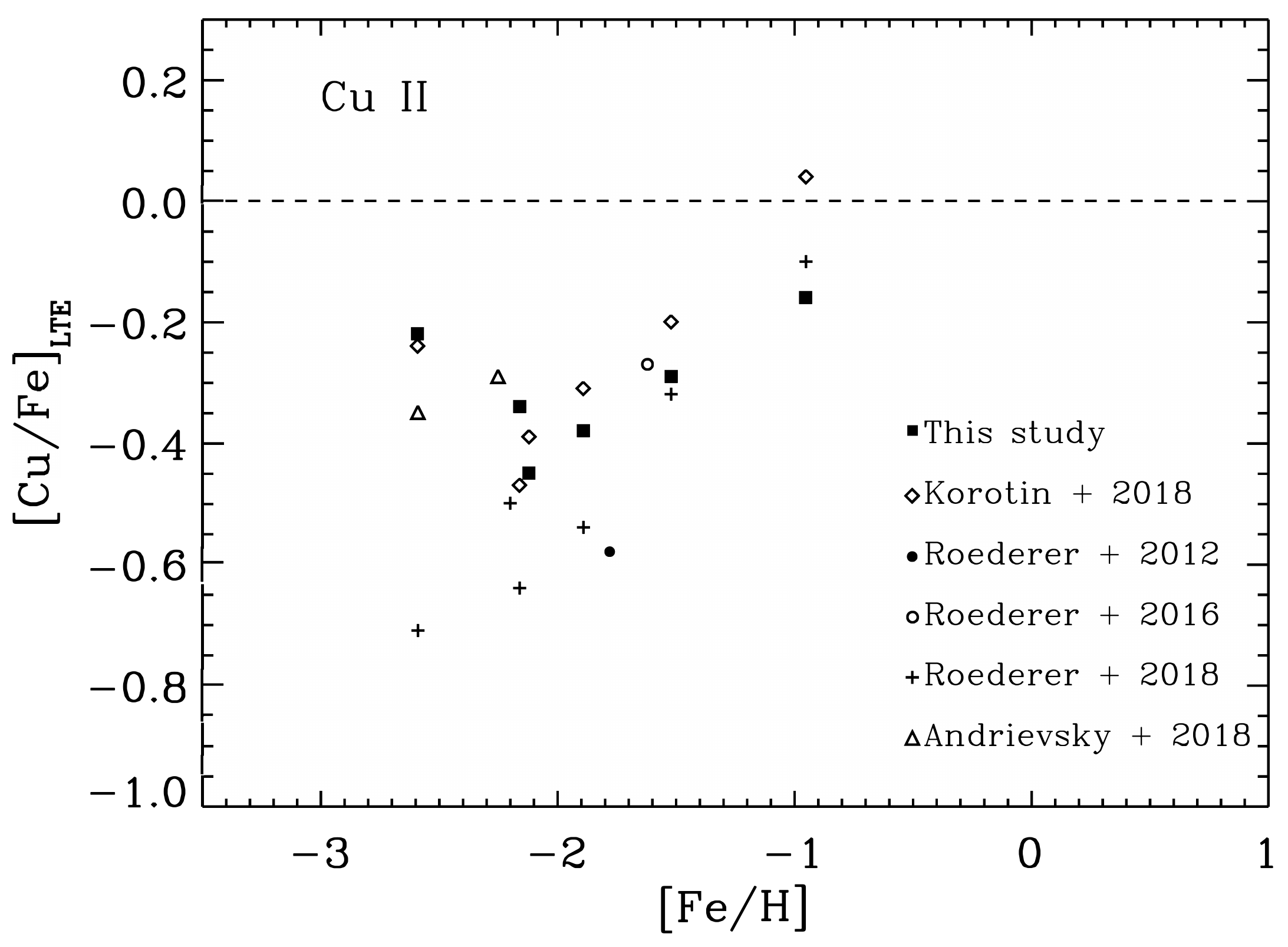}
\caption{Derived Cu abundances. The left panel shows [Cu/Fe] derived from the \ion{Cu}{1} lines for our program stars, which are designed by filled
squares (new model) and open squares (old model). The samples of \cite{2018MNRAS.473.3377A}, \cite{2018MNRAS.480..965K} and \cite{2018ApJ...862...71S} are designed by 
open triangles, diamonds and grey filled circles. The theoretical predictions of \citet[black solid line]{2020ApJ...900..179K} are also presented.
The right panel shows [Cu/Fe] derived from the \ion{Cu}{2} lines for our sample stars. The samples of \cite{2018MNRAS.480..965K} and \cite{2018ApJ...857....2R}, 
two stars analyzed by \cite{2018MNRAS.473.3377A}, one star analyzed by \cite{2012ApJ...750...76R}, and one star analyzed by \cite{2016ApJ...821...37R} are also shown.
Corresponding symbols are annotated in the figure.
\label{fig.trend}} 
\end{figure*}

\subsection{The Galactic [Cu/Fe] Trend} \label{subsec:evolution}

The [Cu/Fe] ratios play an important role in revealing the nature of Cu nucleosynthesis and constraining the Galactic chemical evolution (GCE) model.
In Figure\,\ref{fig.trend} we illustrate the Cu abundances derived in this study. The left panel of Figure\,\ref{fig.trend} compares the NLTE [Cu/Fe] ratios
determined from the \ion{Cu}{1} lines with a representative selection of previous works from the literature. The references are listed in the caption of 
Figure\,\ref{fig.trend}. It can be seen that our NLTE Cu abundances are consistent with the results of \cite{2018ApJ...862...71S}, excluding the group of stars 
with low-$\alpha$ ratios as discussed by \cite{2018ApJ...862...71S}.
Meanwhile, our [Cu/Fe] ratios are systematically lower than those from \cite{2018MNRAS.473.3377A} and \cite{2018MNRAS.480..965K}, and we suggest it could be ascribed
to the difference of adopted copper atomic models as discussed in Section\,\ref{subsec:comparison}. 
Moreover, through the whole metallicity range of our program stars, 
the features of the [Cu/Fe] trend have no large difference between the results when using the old and new models. There is only a downward shift
in the [Cu/Fe] ratios with respect to that of the old model, namely the application of accurate collisional data for
our sample stars leads to a lower NLTE copper abundance.

The right panel of Figure\,\ref{fig.trend} presents the [Cu/Fe] ratios derived from \ion{Cu}{2} lines, and we can see that our LTE Cu abundances are overall in good
agreement with the results in the literature. Although our sample stars are small, a feature of the [Cu/Fe] trend can still be found:
it increases with increasing metallicity when $\sim$\,$-$2.0\,$<$\,[Fe/H]\,$<$\,$\sim$\,$-$1.0\,dex, which indicates that copper behaves as a secondary 
(metallicity-dependent) element. The rise of [Cu/Fe] ratios with increasing metallicity is most likely due to the metallicity-dependent Cu yields from the weak 
\emph{s}-process in massive stars \citep[SN\,II progenitors][]{2004MmSAI..75..741B}. While in the metallicity range of [Fe/H]\,$<$\,$\sim$\,$-$2.0\,dex,
the feature is not clear due to the small sample of stars.

Many groups \citep[e.g.][]{1991A&A...246..354S,1995ApJS...98..617T, 2000A&A...359..191G, 2002A&A...396..189M, 2006ApJ...653.1145K, 2011MNRAS.414.3231K, 2020ApJ...900..179K, 2007MNRAS.378L..59R, 2010A&A...522A..32R}
have attempted to model the Galactic evolution of copper. In Figure\,\ref{fig.trend}, we overplot the behavior of [Cu/Fe] trend predicted by the 
Galactic chemical evolution (GCE) model from \cite{2020ApJ...900..179K}. We note that our [Cu/Fe] ratios are consistent with the model predictions 
in the metallicity range $\sim$ $-$2.0 $<$ [Fe/H] $<$ $\sim$ $-$1.0\,dex. However, when [Fe/H]\,$<$\,$\sim$ $-$2.0\,dex,
the model predictions can not reproduce the [Cu/Fe] ratios. It needs to be pointed out that 
our sample of stars with metallicity less than $\sim$ $-$2.0\,dex, is small,
and more extremely metal-poor stars are needed in future work to explain the discrepancy.

\section{Conclusions} \label{sec:conclusion}

Our work mainly aims to perform the NLTE calculations for \ion{Cu}{1} lines using the data from \cite{2021MNRAS.501.4968B} for inelastic 
collisions with hydrogen. We derived the NLTE and LTE abundances from the \ion{Cu}{1} and \ion{Cu}{2} lines for
six warm, metal-poor stars in the metallicity range of $-$2.59\,$\le$\,[Fe/H]\,$\le$\,$-$0.95. Despite the small sample of stars
studied, we can still draw the following noticeable conclusions: 

\begin{enumerate}
\item Our results show that the NLTE effects are obvious either for optical or ultraviolet \ion{Cu}{1} lines, and confirm that
the NLTE effects have a dependency on the metallicity, increasing with decreasing [Fe/H], found by \cite{2015ApJ...802...36Y}.

\item The application of accurate data leads to a decrease in the departure from LTE and lower NLTE copper abundances for neutral Cu lines.

\item When the uncertainties are taken into account, consistent abundances from the \ion{Cu}{1} (NLTE) and \ion{Cu}{2} (LTE) 
lines are obtained. It suggests that our adopted copper atomic model is valid for investigating the formation of \ion{Cu}{1} lines.

\item The [Cu/Fe] ratios increase with increasing metallicity when $\sim$\,$-$2.0\,$<$\,[Fe/H]\,$<$\,$\sim$\,$-$1.0\,dex, which indicates a secondary (metallicity-dependent)
 production of Cu.
\end{enumerate} 

To the best of our knowledge, this work is the first NLTE investigation of \ion{Cu}{1} lines to explore the influence on NLTE effects
and copper abundances using the theoretical hydrogen collision data. Our results indicate that it is essential
to perform NLTE abundance analysis for both optical and UV \ion{Cu}{1} lines in metal-poor stars.

Our research is partially supported by the Scholar Program of Beijing Academy of Science and Technology (DZ:BS202002). 
We acknowledge the science research grants from the China Manned Space Project with No.\,CMS-CSST-2021-B05. 
This work is also supported by the National Key R\&D Program of China No.\,2019YFA0405502, the National Natural Science
Foundation of China under grant Nos.\,12090040, 12090044, and 11833006.  We acknowledge the data provided by the Mikulski Archive
for Space Telescope (MAST) and the ESO Science Archive.

\bibliography{Cu_Atom_xdxu}{}
\bibliographystyle{aasjournal}

\end{document}